\documentclass[twocolumn,showpacs,preprintnumbers,amsmath,amssymb]{revtex4}
\usepackage{graphicx}

\begin{document}

\title{Tri-layer superlattices: A route to magnetoelectric multiferroics?}

\author{Alison J. Hatt}
\author{Nicola A. Spaldin }
\affiliation{Materials Department, University of California}

\date{\today}

\begin{abstract}
We explore computationally the formation of tri-layer superlattices as
an alternative approach for combining ferroelectricity with magnetism to form magnetoelectric multiferroics. We find that the contribution to the superlattice polarization from tri-layering is small compared to typical polarizations in conventional ferroelectrics, and the switchable ferroelectric component is negligible. In contrast, we show that epitaxial strain and ``negative pressure'' can yield large, switchable polarizations that are compatible with the coexistence of magnetism, even in materials with no active ferroelectric ions. 
\end{abstract}

\maketitle

The simultaneous presence of ferromagnetism and ferroelectricity in
magnetoelectric multiferroics suggests tremendous potential for
innovative device applications and exploration of the fundamental physics
of coupled phenomena.
However, the two properties are chemically {\em contra-indicated},
since the transition metal $d$ electrons which are favorable for 
ferromagnetism disfavor the off-centering of cations required 
for ferroelectricity \cite{Hill:2000}. Continued progress in this 
burgeoning field rests on the identification of alternative mechanisms 
for ferroelectricity which are compatible with the existence of
magnetism \cite{Ederer/Spaldin_COSSMS:2005,Fiebig:2005}. Mechanisms
discovered to date include the incorporation of stereochemically
active lone pair cations, for example in BiMnO$_3$ 
\cite{Seshadri/Hill:2001,dosSantos_et_al_SSC:2002} and BiFeO$_3$ 
\cite{Wang_et_al:2003, Neaton_et_al:2005}, geometric ferroelectricity
in YMnO$_3$ \cite{vanAken_et_al:2004}, BaNiF$_4$ 
\cite{Fox/Scott:1977,Ederer/Spaldin:2006} and related compounds,
charge ordering as in LuFe$_2$O$_4$ 
\cite{Ikeda_et_al:2005,Subramanian_et_al:2006},
and polar magnetic spin-spiral states, of which TbMnO$_3$ is
the prototype \cite{Kimura_et_al_Nature:2003}. However, there are 
currently no single phase multiferroics with large and robust 
magnetization and polarization at or near room temperature \cite{Ramesh/Spaldin:2007}.

The study of ferroelectrics has been invigorated over the last few years
by tremendous improvements in the ability to grow high quality
ferroelectric thin films with precisely controlled composition, atomic
arrangements and interfaces. In particular, the use of compositional
ordering that breaks inversion symmetry, such as the layer-by-layer
growth of three materials in an A-B-C-A-B-C... arrangement, has produced
systems with enhanced polarizations and large non-linear optical
responses
\cite{Sai/Meyer/Vanderbilt:2000,Lee_et_al:2005,Warusawithana_et_al:2003,Ogawa_et_al:2003}.
Here we explore computationally this tri-layering approach as an
alternative route to magnetoelectric multiferroics. Our hypothesis is
that the magnetic ions in such a tri-layer superlattice will be
constrained in a polar, ferroelectric state by the symmetry of the
system, in spite of their natural tendency to remain centrosymmetric. We
note, however, that in previous tri-layering studies, at least one of the
constituents has been a strong ferroelectric in its own right, and the
other constituents have often contained so-called second-order
Jahn-Teller ions such as Ti$^{4+}$, which have a tendency to off-center.
Therefore factors such as electrostatic effects from internal electric
fields originating in the strong ferroelectric layers
\cite{Neaton/Rabe:2003}, or epitaxial strain, which is well established
to enhance or even induce ferroelectric properties in thin films with
second-order Jahn-Teller ions
\cite{Choi_et_al:2004,Wang_et_al:2003,Haeni_et_al:2004}, could have been
responsible for the enhanced polarization in those studies.

\begin{figure}[h]
\includegraphics{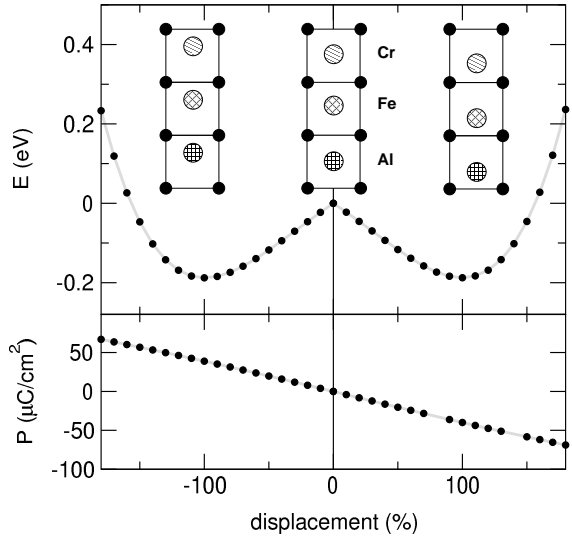}
\caption{\label{doublewell}Energy and polarization as a function of
displacement from the centrosymmetric structure for La(Al,Fe,Cr)O$_3$ under
negative pressure with $a$ = $c/3$ = 3.95 \AA.  Inset: Schematic representation
of the centrosymmetric unit cell (center) and displacements of the metal
cations corresponding to the energy minima.  Displacements are exaggerated for
clarity.}
\end{figure}

We choose a [001] tri-layer superlattice of perovskite-structure
LaAlO$_3$, LaFeO$_3$ and LaCrO$_3$ as our model system (see
Fig.~\ref{doublewell}, inset.)  Our choice is motivated by three
factors. First, all of the ions are filled shell or filled sub-shell,
and therefore insulating behavior, a prerequisite for ferroelectricity,
is likely. Second, the Fe$^{3+}$ and Cr$^{3+}$ will introduce
magnetism.  And third, none of the parent compounds are ferroelectric
or even contain ions that have a tendency towards ferroelectric
distortions, allowing us to test the influence of trilayering alone as
the driving force for ferroelectricity.  For all calculations we use
the LDA+$U$ method \cite{Anisimov/Aryasetiawan/Liechtenstein:1997} of
density functional theory as implemented in the Vienna Ab-initio
Simulation Package (VASP) \cite{Kresse/Furthmuller:1996}. We use the
projector augmented wave (PAW) method
\cite{Bloechl:1994,Kresse/Joubert:1999} with the default VASP
potentials (La, Al, Fe\_pv, Cr\_pv, O), a 6x6x2 Monkhorst-Pack mesh and
a plane-wave energy cutoff of 450 eV.  Polarizations are obtained using
the standard Berry phase technique
\cite{King-Smith/Vanderbilt:1993,Vanderbilt/King-Smith:1993} as
implemented in VASP.  We find that $U/J$ values of 6/0.6 eV and 5/0.5
eV on the Fe and Cr ions respectively, are required to obtain
insulating band structures; smaller values of $U$ lead to metallic
ground states.  These values have been shown to give reasonable
agreement with experimental band gaps and magnetic moments in related
systems \cite{Yang_et_al:1999} but are somewhat lower than values
obtained for trivalent Fe and Cr using a constrained LDA approach
\cite{Solovyev/Hamada/Terakura:1996}. We therefore regard them as a
likely lower limit of physically meaningful $U/J$ values.
(Correspondingly, since increasing $U$ often decreases the covalency of
a system, our calculated polarizations likely provide upper bounds to
the experimentally attainable polarizations). 

\begin{figure}[h]
\includegraphics{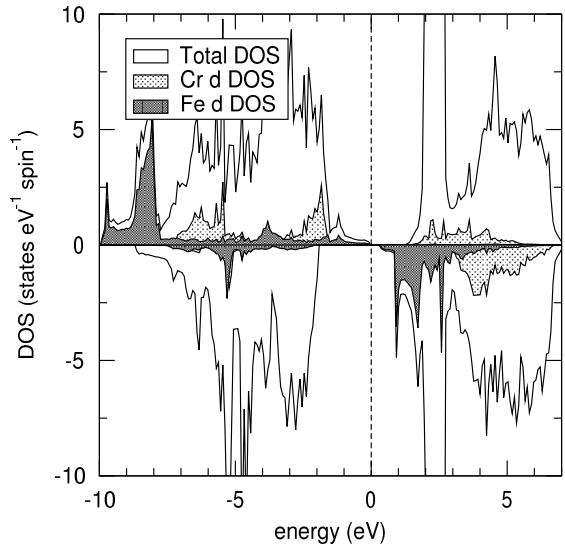}
\caption{\label{dosplot}Density of states for Fe and Cr ions in La(Al,Fe,Cr)O$_3$ with U/J values of 6/0.6 eV and 5/0.5 eV respectively. The dashed line at 0 eV indicates the position of the Fermi energy.}
\end{figure}

We begin by constraining the in-plane $a$ lattice constant to the LDA
lattice constant of cubic SrTiO$_3$ (3.85 \AA) to simulate growth
on a substrate, and adjust the
out-of-plane $c$  lattice constant until the stress is minimized, with
the ions constrained in each layer to the ideal, high-symmetry
perovskite positions. We refer to this as our reference 
structure.  (The LDA (LDA+U) lattice constants for cubic LaAlO$_3$
(LaFeO$_3$, LaCrO$_3$) are 3.75, 3.85 and 3.84 \AA, respectively.
Thus, LaAlO$_3$ is under tensile strain and LaFeO$_3$/LaCrO$_3$ are
unstrained.)  The calculated total density of states, and the local
densities of states on the magnetic ions, are shown in Figure
\ref{dosplot}; a band gap of 0.32 eV is clearly visible.  The
polarization of this reference structure differs from that of the
corresponding non-polar single-component material (for example pure
LaAlO$_3$) at the same lattice parameters by 0.21 $\mu$C/cm$^2$ .
Note, however, that this polarization is not switchable by an electric
field since it is a consequence of the tri-layered arrangement of the
B-site cations.  Next, we remove the constraint on the high symmetry
ionic positions, and relax the ions to their lowest energy positions
along the $c$ axis by minimizing the Hellmann-Feynman forces, while
retaining tetragonal symmetry. We obtain a ground state that is
significantly (0.14 eV) lower in energy than the reference structure,
but which has a similar value of polarization. Two stable ground states
with different and opposite polarizations from the reference structure,
the signature of a ferroelectric, are not obtained.  Thus it appears
that tri-layering alone does not lead to a significant switchable
polarization in the absence of some additional driving force for
ferroelectricity.  In all cases, the magnetic ions are high spin with
negligible energy differences between ferro- and ferri-magnetic
orderings of the Fe and Cr ions; both arrangements lead to substantial
magnetizations of 440 and 110 emu/cm$^3$ respectively. Such magnetic
tri-layer systems could prove useful in non-linear-optical
applications, where a breaking of the inversion center is required, but
a switchable polarization is not.  

\begin{figure}[h]
\includegraphics{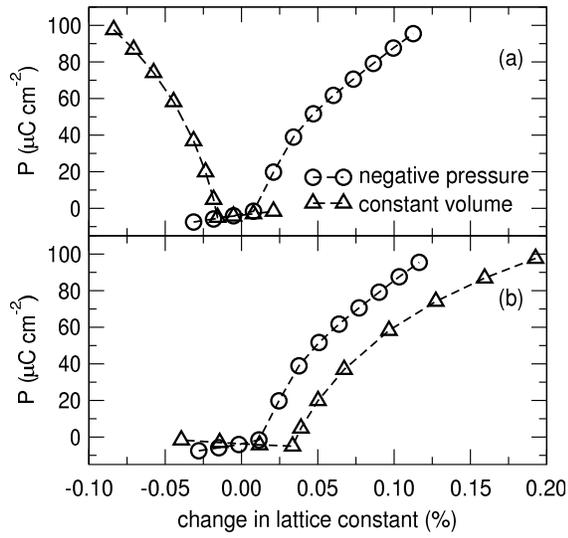}
\caption{\label{phasetrans} Calculated polarizations of negative pressure (circles) and epitaxially strained (triangles) La(Al,Fe,Cr)O$_3$ as a function of change in (a) in-plane and (b) out-of-plane lattice constants relative to the lattice constants of the fully relaxed structures. The polarizations are reported relative to the appropriate corresponding reference structures in each case.}
\end{figure}

Since epitaxial strain has been shown to have a strong influence on the
polarization of some ferroelectrics (such as increasing the remanent
polarization and Curie temperature of BaTiO$_3$ \cite{Choi_et_al:2004}
and inducing room temperature ferroelectricity in otherwise
paraelectric SrTiO$_3$ \cite{Haeni_et_al:2004}) we next explore the
effect of epitaxial strain on the polarization of La(Al,Fe,Cr)O$_3$. To
simulate the effects of epitaxial strain we constrain the value of the
in-plane lattice parameter, adjust the out of plane parameter so as to
maintain a constant cell volume, and relax the atomic positions.  The
volume maintained is that of the calculated fully optimized structure,
167 \AA$^3$, which has an in-plane lattice constant of 3.82 \AA.  As
shown in Figure \ref{phasetrans}, we find that La(Al,Fe,Cr)O$_3$
undergoes a phase transition to a polar state at an in-plane lattice
constant of 3.76 \AA, which corresponds to a (compressive) strain of
-0.016 (calculated from $(a_{\parallel}-a_0)/a_0$ where a$_{\|}$ is the
in-plane lattice constant and a$_0$ is the calculated equilibrium
lattice constant).  A compressive strain of -0.016 is within the range
attainable by growing a thin film on a substrate with a suitably
reduced lattice constant.  

We find that significant ferroelectric polarizations
can be induced in La(Al,Fe,Cr)O$_3$ at even smaller strain values by
using {\it negative pressure} conditions. We simulate negative
pressure by increasing all three lattice constants and 
imposing the constraint a=b=c/3; such a growth condition might be realized
experimentally by growing the film in small cavities on the surface
of a large-lattice-constant substrate, such that epitaxy occurs both
horizontally and vertically.  As in the planar epitaxial strain
state, the system becomes strongly polar; this time the phase
transition to the polar state occurs at a lattice constant of 3.85
\AA, at which the strain is a negligible 0.001 relative to the
lattice constant of the fully optimized system.  

In Fig.~\ref{doublewell} we show the calculated energy versus
distortion profile and polarization for negative pressure
La(Al,Fe,Cr)O$_3$ with in-plane lattice constant = 3.95 \AA, well within the
ferroelectric region of the phase diagram shown in
Fig.~\ref{phasetrans}. The system has a
characteristic ferroelectric double well potential which is almost
symmetric in spite of the tri-layering; the
two ground states have polarizations of 38.9 and -39.9
$\mu$C cm$^{-2}$ respectively, relative to the reference
structure at the same lattice constant. Since the energies of the
two minima are almost identical, the effective electric field
E$_{eff}$=$\Delta$E/$\Delta$P, introduced in Ref
\cite{Sai/Meyer/Vanderbilt:2000}, is close to zero and
there is no tendency to self pole.  The origin of the symmetry is
seen in the calculated Born effective charges (3.6, 3.5 and 
3.3 for Al, Fe and Cr respectively) which show that the system
is largely ionic, with the ions showing
very similar trivalent cationic behavior. A similar profile is
observed under planar epitaxial strain, although the planar strained
system is around 0.15 eV lower in energy than the negative pressure
system for the same in-plane lattice constant.  

To decouple the
effects of interfacial strain and tri-layering we calculate the
polarization as a function of strain and negative pressure for the
individual components, LaAlO$_3$, LaFeO$_3$ and LaCrO$_3$. We find
that all three single-phase materials become polar at 
planar epitaxial strains of -0.03 (LaAlO$_3$),
-0.02 (LaFeO$_3$), and -0.01 (LaCrO$_3$).  Likewise, all three
components become polar at negative pressure, under strains of +0.03
(LaAlO$_3$), +0.001 (LaFeO$_3$), and +0.001 (LaCrO$_3$). (The
higher strains required in LaAlO$_3$ reflect its smaller
equilibrium lattice constant.) 

These results confirm our earlier conclusion that the large polarizations obtained
in strained and negative pressure La(Al,Fe,Cr)O$_3$ are not a result
of the tri-layering.  We therefore suggest that many perovskite
oxides should be expected to show ferroelectricity provided that two
conditions imposed in our calculations are met: First, the ionic
radii of the cation sites in the high symmetry structure are larger
than the ideal radii, so that structural distortions are desirable
in order to achieve an optimal bonding configuration.  This can be
achieved by straining the system epitaxially or in a ``negative
pressure'' configuration. And second, non-polar structural
distortions, such as Glazer tiltings \cite{Glazer:1972}, are
de-activated relative to polar, off-centering distortions.  These
have been prohibited in our calculations by the imposition of
tetragonal symmetry; we propose that the symmetry constraints
provided experimentally by hetero-epitaxy in two or three dimensions
should also disfavor non-polar tilting and rotational distortions. A
recent intriguing theoretical prediction that disorder can be used
to disfavor cooperative tilting modes is awaiting experimental
verification \cite{Bilc/Singh:2006}.

Finally, we compare the tri-layered La(Al,Fe,Cr)O$_3$ with
the polarization of its individual components. Calculated
separately, the remnant polarizations of LaAlO$_3$, LaFeO$_3$ and
LaCrO$_3$, all at negative pressure with a=c=3.95 \AA, average to 40.4
$\mu$C cm$^{-2}$.  This is only slightly larger than the calculated
polarizations of the heterostructure, 38.9 and 39.9 $\mu$C cm$^{-2}$,
indicating that tri-laying has a negligible effect on the polarity. 
This surprizing result warrants further investigation
into how the layering geometry modifies the overall polarization.  

In conclusion, we have shown that asymmetric layering alone is not
sufficient to produce a significant switchable polarization in a
La(Al,Fe,Cr)O$_3$ superlattice, and we suggest that earlier
reports of large polarizations in other tri-layer structures may have resulted 
from the intrinsic polarization of one of the components combined
with epitaxial strain. We find instead that La(Al,Fe,Cr)O$_3$
and its parent compounds can
become strongly polar under reasonable values of epitaxial strain
and symmetry constraints, and that tri-layering serves to modify the
resulting polarization. Finally, we suggest ``negative pressure''
as an alternative route to ferroelectricity and hope that our prediction motivates 
experimental exploration of such growth techniques.   

This work was funded by the NSF IGERT program, grant number
DGE-9987618, and the NSF Division of Materials Research, grant number
DMR-0605852.
The authors thank Massimiliano Stengel and Claude Ederer
for helpful discussions.

\bibliography{alison}

\end{document}